\documentclass[aps,pra,onecolumn]{revtex4}
\usepackage{graphicx}
\begin{document}

\title{Conflicts Between Science and Religion: Epistemology to the Rescue}

\author{Moorad Alexanian}

\affiliation{Department of Physics and Physical Oceanography\\
University of North Carolina Wilmington\\ Wilmington, NC
28403-5606\\}
\date{\today}

\begin{abstract} Both Albert Einstein and Erwin Schr\"{o}dinger have defined what science is. Einstein includes not only physics, but also all natural sciences dealing with both organic and inorganic processes in his definition of science.  According to Schr\"{o}dinger, the present scientific worldview is based on the two basic attitudes of comprehensibility and objectivation. On the other hand, the notion of religion is quite equivocal and unless clearly defined will easily lead to all sorts of misunderstandings.  Does science, as defined, encompass the whole of reality? More importantly, what is the whole of reality and how do we obtain data for it? The Christian worldview considers a human as body, mind, and spirit (soul), which is consistent with Cartesian ontology of only three elements: matter, mind, and God. Therefore, is it possible to give a precise definition of science showing that the conflicts are actually apparent and not real?\\
\end{abstract}

\maketitle {}

\section{Introduction}
In 1950, Albert Einstein gave a remarkable lecture to the International Congress of Surgeons in Cleveland, Ohio.  Einstein argued that the 19th-century physicists' simplistic view of Nature gave biologists the confidence to treat life as a purely physical phenomenon.  This mechanistic picture of Nature was based on the casual laws of Newtonian mechanics and the Faraday-Maxwell theory of electromagnetism.  These causal laws proved to be wanting, especially in atomistic phenomena, which brought about the advent of quantum mechanics in the 20th-century. Einstein indicates that there are three principal features \cite{ET05} which science has firmly adhered to since Galileo Galilei.

In the process of studying the whole of reality, humans collect data that is subsequently analyzed according to the different kinds of knowledge. The study of the different aspects of a given element of reality gives rise to different kinds of knowledge, which leads to a need to classify kinds of knowledge as either autonomous or synthetic whose integration via a supposed metaphysics gives rise to a composite nature of all elements of reality. The different kinds of knowledge are characterized by their subject matters and deal primarily with a certain aspect of reality and thus determines what kind of evidence is necessary to establish the truth or falsehood of a given proposition in that particular kind of knowledge. One of the crucial problems in epistemology is the nature and possible existence of a demarcation of science from non-science, which is essential for the resolution of the apparent conflicts between science and religion. It is foolish for a scientist to require the same kind of evidence, which is appropriate to establish truthful statements in science, from a theologian, which has its own source of evidentiary data. It may be that the elements of reality studied by both the scientist and the theologian are the same; however, a scientist studies only the purely physical aspect of that entity, whereas a theologian is concerned mainly with the nonphysical and possible supernatural nature. In the set-theoretic description of the whole of reality \cite{MA06}, a human being possesses both physical and nonphysical aspects, with the physical being the concern of the scientist whereas the nonphysical, say, life, consciousness and rationality, may actually represent manifestations of the supernatural. Jacques Maritain indicated the necessity of this sort of Cartesian dualism \cite{JM61}. Herewith, the ``phenomenally" of Maritain is the physical data of science, whereas his ``ontologically" is that ``data" collected by human being as ``detectors" of that aspect of reality that escapes the purely physical devices.  Of course, one ought never to forget that human rationality characterizes the whole of reality by nonphysical mental models, abstractions, and constructs that have their counterparts in the real but are not identical to them.

The nonphysical aspect of consciousness and rationality together with the physical aspect of humans gives rise to the traditional mind-body problem that poses the following dilemma, ``First: if the mind is not physical, then how can it have effects in the physical world? But, second, if the mind is physical, then how can we understand consciousness? The first question drives us towards physicalism, while the second drives us towards dualism." \cite{CF04} It may be that the supernatural in humans actually mediates between the nonphysical mind and the physical body and, similarly, exercises the free will in our actions \cite{MA08}. This assertion is based on the observation that the Apostle Peter needed the Father in order to know the true nature of Jesus (Matt. 16:16-17).  Therefore, the supernatural in humans is the ``detector" and the seat of self as well as the means to know or ``detect" God.   This is somewhat reminiscent of the doctrine of occasionalism of Nicolas Malebranche where God is the only true and active causal agent in the universe \cite{SN00a}. The inefficacy of natural causes regarding relations between bodies also applies to the mind-body interaction and so God himself mediates between the human mind and body \cite{SN97}. One may consider Malebranche's proposal as a realization of God upholding the creation moment-by-moment (Hebrews 1:3).  It is interesting that Malebranche considers individual human minds as limitations of the universal mind of God, which is somewhat similar to Schr\"{o}dinger's consideration of the Vedantic notion that all consciousness is essentially one, viz., the oneness of mind \cite{ES67}.

\section{Metaphysics: E. Schr\"{o}dinger and C. S. Lewis}
Erwin Schr\"{o}dinger was the forerunner in evolutionary biology, genetics, and indeed a great philosopher \cite{JG92}. P.A.M. Dirac remarked that Schr\"{o}dinger's equation underlies ``a large part of physics and the whole of chemistry" \cite{PAMD29}.  Schr\"{o}dinger was puzzled by the agreement of the existence of a common, real world observed by two differing observers. ``Each person's sense-world is strictly private and not directly accessible to anyone else, this agreement is strange, what is especially stranger is how it is established" \cite{ES83a}. Schr\"{o}dinger asked, ``How do we come to know of this general agreement between two private worlds, when they admittedly are private and always remain so?" \cite{ES83b} Concerning his holistic view of Nature, Schr\"{o}dinger considers two hypotheses \cite{ES83c}. Schr\"{o}dinger further indicates, ``I have therefore no hesitation in declaring quite bluntly that the acceptance of a really existing material world, as the explanation of the fact that we all find in the end that we are empirically in the same environment, is mystical and metaphysical"\cite{ES83d}. Schr\"{o}dinger vehemently rejects the idea of an individual soul dwelling in each human but considers instead ``that the external world and consciousness are one and the same thing, in so far as both are constituted by the same primitive elements"\cite{ES83e}.  Also, ``There is no plurality here whatever" \cite{ES83f}. Schr\"{o}dinger's Doctrine of Identity, that despite the diversity of human bodies there is but only one subjective single Self, represents this Self as being unchanging and eternal.  Now consciousness is a moment-by-moment awareness of our temporal existence and surroundings.  Thus, human knowledge has access only to snapshots and flashbacks of reality. In Christian theology, God is the being forever conscious and thus eternal that does not exist in time. God has no history and so He experiences the whole of reality as an eternal ``now" \cite{CSL78}.

The metaphysical and mystical nature that Schr\"{o}dinger ascribes to different spheres of consciousness that recognize that we all live in the same world can partially be demystified if one considers the objective nature of scientific data that is collected with purely physical devices. Of course, the metaphysical and mystical nature of consciousness, rationality and even life itself remains.

C.S. Lewis seeks answers to the same fundamental questions regarding the acquisition of knowledge by humans.  In the process of studying the notion of miracles, their possible occurrence, and their supernatural nature, Lewis makes a detailed analysis of what Nature is and what is the nature of human rationality and morality. ``If our argument has been sound, rational thought or Reason is not interlocked with the great interlocking system of irrational events which we call Nature" \cite{CSL71a}. In particular, ``Hence every theory of the universe which makes the human mind a result of irrational causes is inadmissible, for it would be a proof that there are no such things as proofs. Which is nonsense" \cite{CSL71b}. The laws of Nature govern the physical aspect of Nature and thus possess no elements of free will or rationality. It is in this sense, that the behavior of Nature is understood by Lewis to be irrational.  Of course, there may be intelligence behind the workings of Nature, as in the Christian faith where God not only created the whole of reality but He also sustains the creation moment by moment into a continuous state of existence.

Schr\"{o}dinger posits the need of the metaphysical and the mystical, whereas Lewis considers rationality as necessary elements to understand natural, spatiotemporal events. These are actually the assumptions of comprehensibility and objectivation considered by Schr\"{o}dinger \cite{ES56a}. Lewis further indicates human reason and morality as proofs of the supernatural \cite{CSL71c}.

Conscious, rational humans develop the theories of the workings of the physical aspect of Nature. However, human rationality is not present in the physical but forms part of the nonphysical aspect of humans---self, consciousness, and rationality. It is inherently the differing theological presuppositions of Schr\"{o}dinger and of Lewis that the higher order inferences of Schr\"{o}dinger ends with the Upanishads and that of Lewis with the Bible. In particular, ``why our perceiving and thinking self is nowhere to be found within the world-picture: because it itself is this world-picture" \cite{ES56b}. Schr\"{o}dinger does not ascribe individuality or distinctness to self whereas Lewis does \cite{CSL71d}.

The relationship of human consciousness, rationality, self, and free will vis-\`{a}-vis the external, spatiotemporal world may be viewed as follows. The cosmological principle, which assumes that the universe is homogenous and isotropic, leads to a model whereby the dynamical Einstein-Hilbert equations determine the behavior of a substratum, a mass average of the existing galaxies, which allows us to deduce observable properties of the physical universe \cite{HB68a}. In addition, it is assumed that ``the velocities of matter in each astronomical neighbourhood (each group of galaxies) are small" \cite{HB68b}. In our analogy, the cosmological substratum represents interlocked Nature where physical laws determine its time development. The local galaxies, with their corresponding small velocities with respect to the substratum, represent the selves in humans, which corresponds to the nonphysical and supernatural aspects of humans, whose actions are determined by the exercise of their free will. The laws of Nature inexorably govern locally the physical aspect of humans and the overall or global development of the Universe; however, human rationality and consciousness govern the exercise of human free will. The latter are not interlocked with the physical aspect of Nature but represent the supernatural elements in Nature \cite{ES56c}.  Of course, the spatial-temporal behavior of the physical universe may not be truly autonomous.  In Christian theology, the Son ``upholds all things by the word of His power" (Hebrews 1:3); however, one truly does not know how His actions manifest themselves in the physical aspect of reality. It may be that the interaction of the Creator with His creation is suggestive of the elusive interaction between the human body and the mind, which would be mediated by the supernatural in humans. This is indeed the ``doctrine of occasional causes," as Leibniz labeled Malebranche's occasionalism \cite{SN00b}.

\section{Nature of Human Knowledge}
One may say that the whole of reality consists of all the data that is known to humans both of the immediate present and past occurrences. I hesitate to say that the whole of reality is equivalent to all that exists and has existed since that would presuppose stronger ontological assumptions and considerably more knowledge of all that is and was that I am supposing here. In fact, our earthly environment consists essentially of living or dead bodies of plants and animals. Schr\"{o}dinger considers the relation of organic to inorganic with ``inorganic matter---the subject-matter, by definition, of physics and chemistry---is an abstraction which, unless by especial arrangement, we actually encounter scarcely anywhere, or at any rate extremely seldom" \cite{ES83g}. Schr\"{o}dinger ascribes the meaning of organic with the criterion of metabolism and quotes Arthur Schopenhauer regarding the demarcation between organic and inorganic beings \cite{ES83h}. Defining the subject matter of science as the physical aspect of Nature is quite consistent with Schopenhauer's demarcation since the inorganic is the physical and the organic the living owing to the metabolism in living organisms that maintains life. Organic beings represent elements in Nature that are both physical and nonphysical since life cannot be reduced to the purely physical whereas; inorganic entities are purely physical entities.

The question of what really exists and how that relates to human sense observations and detection by purely physical devices is a deep philosophical/metaphysical problem that can only be partly addressed by making some basic presuppositions.  Schr\"{o}dinger's views regarding consciousness, our soul, the self and the meaning of the subjective ``I" is not that based on the traditional Christian thought but on Schopenhauer's idea that our real essence is will, which is the ``thing-in-itself," and the doctrine of the Upanishads \cite{ES83j,WL92}. Schr\"{o}dinger believed that all consciousness is essentially one and that a purely rational worldview is an absurdity.

Humans describe the workings of Nature and understand the totality of the human experience with the aid of fundamental kinds of knowledge.  Jacques Maritain \cite{JM52} considers two basic questions: the intrinsic diversity of human knowledge and the inner value or nature of knowledge, that is to say, knowledge that is rational and speculative, philosophical and scientific. Maritain \cite{JM62} also classifies different kinds of knowledge by the object or subject matter that the science deals with. However, Maritain \cite{JM62} characterizes every different kind of knowledge by the term ``science" and philosophy or metaphysics as ``the highest of the human sciences, that is, of the sciences which know things by the natural light of reason." For instance, Maritain \cite{JM62} defines theology as ``the science or knowledge of God which we attain naturally by the unassisted powers of reason."  However, Maritain's choice of the term ``science" is too equivocal since it applies to all sorts of knowledge, which is not in accord with the modern sense of the word ``science."  William Oliver Martin \cite{WOM57} follows also Aristotle and St. Thomas Aquinas in considering that different ways of knowing gives us different sciences.  Alexanian \cite{MA07} has adopted the definition of science not only as the study of the physical aspect of Nature but with the further proviso that the data that makes up the subject matter of science is that which can be obtained, in principle, solely by purely physical devices. In fact, the term ``scientist" was introduced in the nineteenth century when natural philosophy became a synonym for physics and science \cite{EG07}. In the words of Schr\"{o}dinger, ``The strange fact that on the one hand all our knowledge about the world around us, both that gained in everyday life and that revealed by the most carefully planned and painstaking laboratory experiments, rests entirely on immediate sense perception, while on the other hand this knowledge fails to reveal the relations of the sense perceptions to the outside world, so that in the picture or model we form of the outside world, guided by our scientific discoveries, all sensual qualities are absent" \cite{ES67a}. It is interesting that Democritus of Abdera understood this state of affair already in the fifth century B.C. prior to the advent of the sophisticated instrumentations of today \cite{ES67b}.

Maritain correctly criticizes scientists who ``cling to the idea that the only object capable of giving rise to an exact and demonstrable knowledge is that which is sense-perceivable and can be subjected to methods of experimental and mathematical analysis ... and they continue to exclude philosophy or to regard it as a sort of mythology which is only fit to satisfy emotional needs" \cite{JM52a}. A more important observation by Maritain and one that is applicable not only to the study of the physical aspect of Nature but also to the whole of reality is, ``A scientific definition does not tell us what a thing is, but only in what way we can agree on the observations and measurements we have taken from nature, so as to get a knowledge, not of the essence of that thing, but merely of the manner in which the signs which refer to its impact on experience and to the modes of verification grouped under its name, can give rise to a coherent language.  If I say `matter,' to the physicist, this word does not denote a substance or a substantial principle whose nature he tried to reveal to us. It merely denotes a system of mathematical symbols built by microphysics upon an immense body of data of observation and measurement, which are furthermore subject to continual revision" \cite{JM52b}.

In the final analysis, all human knowledge is based on mental abstractions and constructions that encode data obtained by not only our subjective senses but, more importantly for science, also objective data obtained with the aid of purely physical devices. Therefore, in essence, there are only two sources of knowledge, sense and revelation from God. Schr\"{o}dinger wrote eloquently and aptly regarding the hypothesis of the real world, ``The world is a construct of our sensations, perceptions, and memories" \cite{ES67c}. Human communication, whether written, visual, auditory, or tactile, is based on an agreed construct used to describe the whole of reality with the information content in mathematical language having the highest logical and symbolic structure.  The issue regarding how our mental construction of reality is faithful to the actual existing entities is a deep philosophical/metaphysical problem that we are not considering here. Therefore, our reasoning and theories are based solely on the information content of data, which is detected by humans and/or purely physics devices. In fact, Schr\"{o}dinger indicates, ``the real is only the complex of sense impressions, all the rest are only pictures" \cite{WJM89}. The set-theoretic description of the whole of reality in terms of the physical, the nonphysical, and the supernatural \cite{MA06}, suggests kinds of knowledge that are autonomous and appropriate for the study of the different realms of reality.  The whole of reality is not only that which is observed, detected, and studied presently, but includes also all the information of past events that is known whether in written form or mentally known by extant, conscious and rational beings. In the first place, the physical aspect of reality is clearly accessible to purely physical devices via purely physical interactions \cite{MA07}. The physical is essentially the subject matter of physics. The realm of the nonphysical is clearly more complicated since it includes living beings, where rationality, consciousness are not physically detectable, and information or knowledge that such beings can and have access to. This information includes knowledge derived from the experimental sciences, history, logic, mathematics, metaphysics, and theology. Whereas purely physical devices can detect the physical aspect of reality, such devices cannot access the nonphysical information content of physical/nonphysical elements of reality.

For instance, a book is a physical/nonphysical element of reality with the physical being the paper and ink that makes up the book while the nonphysical is the information content of the text. Surely, purely physical devices cannot decode and comprehend the encoded information of the text. This applies especially to computers, which are often compared to humans in artificial intelligence studies where the intellectual and even creative differences are thought considerably to narrow down.  Of course, the fundamental difference between computers and human beings is the existence of consciousness in humans that decodes and understands or knows the nonphysical information content of the human brain. That is to say, one may employ robots to collect the purely physical data that forms the subject matter of science. However, that information, which is purely physical, does not lead to knowledge of the physical aspect of reality without the aid of conscious beings.

Humans, books, and computers are physical/nonphysical entities.  However, the difference in humans is the added nonphysical elements of consciousness and rationality, which do not exist in either books or computers. Whereas there is no interaction or any causal connection between the physical and the nonphysical aspects in books or computers, such is not the case in a human being where the nonphysical can interact with the physical via a nonphysical interaction having a seat in the conscious mind that accesses the information content stored in the physical human brain. This interaction between the physical and the nonphysical is the very same interaction that allows a human to develop mental constructs and abstractions from sense observation in the same fashion as when knowledge is transferred from a book to a human while reading the physical text of the book. One must distinguish between information, which is physical \cite{RL91}, and knowledge, which is nonphysical. The former can be quantified physically, whereas the latter needs a conscious, rational mind in order to decipher the nonphysical content in the physical information into nonphysical knowledge, which is the process of knowing or understanding in humans. Claude Shannon founded information theory, which does not deal with information itself but rather with the quantification, the compression, and the communication of data.

Human consciousness and reasoning summarize all physical data into laws and create the mathematical theories that lead to predictions. However, the human element that creates the theories is totally absent from the laws and theories themselves. Accordingly, human consciousness and rationality are outside the bounds of science since they cannot be detected by purely physical devices and can only be ``detected" by the self in humans.  Schr\"{o}dinger goes further along these lines \cite{ES67d}. That is to say; there is an interaction between the physical and the nonphysical mediated by consciousness.  This understanding of the nature of knowledge is quite similar to that expounded by Maritain \cite{JM52c}.

\section{Order and Integration of Knowledge}
The finite nature of the human mind is evident by the need to understand reality by a process of analysis.  This process of taking things apart has resulted in a multitude of disciplines as manifested in the existence of many departments in our institutions of higher learning.  It is clear that each kind of knowledge deals primarily with a certain aspect of reality and as such it is based on a specific type of evidence that are used to establish the truth or falsehood of given propositions in that field.  For instance, it is foolish for a scientist to require the same kind of evidence, which is appropriate to establish truthful statements in the experimental sciences that together with history forms the two domains of the phenomenological, from a theologian who studies the intrinsic nature of God and how He interacts with His creation and has its own source of evidentiary data.  It is important to remark that the data used in all kinds of knowledge is made up of unique historical events represented by historical propositions, which are based on sensations, perceptions, memories, and extant records of past events.  In evolutionary theory, natural selection and genetic drift are the two major mechanisms that produce changes in existing living organisms. Similarly, human reasoning using essentially historical data stored in the brain produces human knowledge and understanding.  It is interesting that C.S. Lewis indicates, ``The rational and moral element in each human mind is a point of force from the Supernatural working its way into Nature" \cite{CSL71}. Therefore, human reasoning is not entirely interlocked with the physical aspect of Nature. This is equivalent to the statement of Rene Descartes that ``the essence of matter is extension and that extended things cannot think" \cite{MDW05}. Descartes based this argument on clear and distinct intellectual perceptions of the essences of mind and matter, not on the fact that he could doubt the existence of one or the other.

What are the basic, autonomous kinds of knowledge needed to analyze and comprehend the whole of reality?   Martin \cite{WOM57} considers as autonomous the following kinds of knowledge:  history (H), metaphysics (M), theology (T), formal logic (FL), mathematics (M), and experimental science (G), with metaphysics and theology constituting the two domains of the ontological context and the others, viz., H, FL, M, and G, as positive kinds of knowledge. History and experimental science are the two domains of the phenomenological context whereas formal logic is the domain of intentional context and mathematics that of formal context.  In addition, Martin \cite{WOM57} considers synthetic kinds of knowledge those that result from the integration of a positive kind of knowledge with the ontological (metaphysics and/or theology). For instance, the human-social sciences, notably, anthropology, culturology, economics, philology, psychology, and sociology are all examples of synthetic kinds of knowledge that integrates scientific studies with the humanities, which fundamentally deals with the nature of humans. Similarly, the philosophy of history, of mathematics, of science, or of Nature, which are synthetic kinds of knowledge, the mode or aspect studied is integrated with the mode of existence or being, which is the subject matter of metaphysics and theology. In a sense, Martin systemized the epistemology and metaphysics of Maritain who dealt with the intrinsic diversity and nature of knowledge, which is rational and speculative, philosophical and scientific.

Martin makes clear that experimental science is positive science, but it is not the same as positivism \cite{WOM57a}. Therefore, one must distinguish between the causes supposed in the model descriptions of different aspects of reality and the actual causes. The former may be described as secondary while the latter as primary, e.g., God being the primary cause as proposed by Malebranche \cite{SN00a}. This may suggest an understanding of Descartes' notions of `universal and primary cause' and `secondary and particular causes' \cite{TMS08}. God is the primary cause of the actual, moment-by-moment temporal development of all that is; whereas, the secondary causes are those that we ascribe to the models that we construct of all that there is, which are based on our sensations, perceptions, and memories.  Our understanding of Descartes' primary causes is in agreement with the `occasionalism' of Malebranche and is contrary to the view of Schmaltz who considers creatures rather than God as the casual source of natural change rather than merely `occasional causes' \cite{TMS08a}.

Knowledge is summarized in propositions. For instance, historical propositions refer to that which is factually known (potentially or actually) of the process of historical events both human and natural past events. Historical propositions are instrumental in a general sense to propositions of metaphysics, formal logic, and mathematics \cite{WOM57b}. Scientific laws of Nature are generalization of historical propositions, the experimental sciences G, where the data expressed in historical propositions can be obtained, in principle, by purely physical devices. Therefore, history H in the form of historical propositions is wholly constitutive of experimental science G.  The subject matter of formal logic FL and mathematics M are the nonphysical, mental or mathematical constructs of the ``real things," which are based on data detected by human senses and/or purely physical devices. Human rationality develops formal logic and creates mathematics to summarize data into laws of Nature that lead to theoretical models covering a wide range of phenomena. However, scientists deal with secondary causes. First causes involve metaphysical (ontological) questions, which regulate science. Without the ontological, neither the generalizations nor the historical propositions of the experimental sciences would be possible.

The integration of all the positive sciences with the ontological gives us the ``real thing" that actually exists. Metaphysics and theology, the fields that encompass the knowledge of being, constitute the two domains of the ontological context that deal with the mode or aspect including the existence or being of the ``real thing."  There is a nested sequence of mental abstractions and constructions in the human mind organizing and making sense of the reality based on physical data obtained by purely physical devices and data obtained by humans as ``living detectors" of the physical, nonphysical, and the supernatural aspects of Nature

In Table 1, the relations between the different autonomous or irreducible kinds of knowledge are specified in terms of, ``constitutive of" (Con), ``instrumental to" (Inst), and ``regulative of" (Reg).  For instance, metaphysics M is regulative of all the different kinds of knowledge, except theology T where metaphysics M is constitutive of some theological propositions. The latter means that some theological proposition $t_{1}$ implies a particular metaphysical proposition $m_{1}$, viz., ``If $t_{1}$, then  $m_{1}$."  The converse, ``If $m_{1}$, then $t_{1}$," does not follow since otherwise the theological proposition $t_{1}$ is constitutive of the metaphysical proposition $m_{1}$ and so the truth of $t_{1}$ would be necessary for the truth of $m_{1}$. The constitutive aspect of historical propositions for the experimental sciences means that given the set of historical propositions $\{h_{i}\}$, $(i = 1, 2,\cdots, n)$ and the generalization of them by $g_{1}$, one has that $g_{1} \rightarrow (h_{1}, h_{2},\cdots, h_{n})$.  Therefore, the set $\{h_{i}\}$ is wholly constitutive of $g_{1}$.  The induction or inference, if $(h_{1}, h_{2},\cdots, h_{n})\rightarrow \textup{(probably)} g_{1}$, is the modus operandi of the experimental sciences where probability theory is used  to indicate which of a given set of generalizations ${g_{1}, g_{2},\cdots}$ is most likely to be true in the light of the data and any other evidence at hand.

Mathematical propositions are instrumental in discovering and summarizing the generalizations, which are constituted by the historical propositions.  Therefore, the generalization $g_{1}$, sometimes codified in a theoretical model, gives rise to predictions, say $g_{1} \rightarrow h_{p}$. If $h_{p}$ is false, then $g_{1}$ is also false and so the generalization or underlying theory is falsified. ``Facts (historical propositions) are as relevant to metaphysics as to experimental science, but not in the same way; for they are instrumental to the discovery of metaphysical truth, but are constitutive as evidence of the generalizations of experimental science" \cite{WOM57c}.

\begin{center}
\begin{tabular}{|p{1.5cm}|p{1.5cm}|p{1.5cm}|p{1.5cm}|p{1.5cm}|p{1.5cm}|p{1.5cm}| }
 \hline
 \multicolumn{7}{|c|}{Table 1: \textbf{Relations between autonomous kinds of knowledge*}  } \\
 \hline
 \hline
Subject  & H & Meta&T&FL&Math&G\\
 \hline
 H&X&Inst&Con of Some&Inst&Inst&Con\\
 \hline
 Meta&Reg&X&Con of Some&Reg&Reg&Reg\\
 \hline
T&None&None&X&None&None&None\\
\hline
 FL&Inst&Inst&Inst&X&Inst&Inst\\
 \hline
 Math&Inst&Inst&Inst&Inst&X&Inst\\
 \hline
 G&Con of Some&Inst&Inst&Inst&Inst&X\\
 \hline
\end{tabular}
\end{center}

*In Martin \cite{WOM57}, the order of knowledge of  historical propositions (H), metaphysical propositions (Meta), theological propositions (T), formal logic propositions (FL), mathematical propositions (M), and the generalizations of experimental science (G) together with their interrelationships described by ``instrumental to" (Inst), ``regulative of" (Reg), and ``constitutive of "(Con).\\

The analysis of elements of reality into its different aspects, viz., physical, nonphysical, or supernatural, gives rise to the different kinds of knowledge needed to give a true description and understanding of that which is detectable by purely physical devices and by humans as ``detectors."  The integration of all kinds of knowledge is the object of metaphysics, which delimits the possible and is regulative of all the positive kinds of knowledge, viz., H, FL, M, and G, and is partially constitutive of T (See Table 1).  Historical propositions dealing with the physical aspect of Nature are constitutive of the generalizations of experimental science and form the basis for unadulterated science and the discovery of the laws of Nature. These generalizations of physical data into laws, say $(g_{1}, g_{2},\cdots g_{10})$, in turn imply a minimal metaphysics $m_{1}$ dictated by some sort of Ockham's razor, which forms a foundation of our understanding of the physical aspect of Nature. Symbolically, one has $(g_{1}, g_{2},\cdots g_{10})  \rightarrow  m_{1}$, where metaphysics $m_{1}$ deals merely with the physical aspect of Nature.  An example of an implied metaphysics is the principle that Nature can be understood and the principle of objectivation \cite{ES67e}. It is clear that any metaphysics $m_{2}$, which contains metaphysics $m_{1}$ as a subset, is equally compatible with the generalizations $(g_{1}, g_{2},\cdots g_{10})$.  The stronger constitutive relation of metaphysics to some theologies means that given theology $t_{1}$, then metaphysics $m_{3}$ must be true, viz., $t_{1} \rightarrow m_{3}$. Theology and metaphysics constitute the two domains of the ontological context of the whole of reality.  Since experimental science G is concerned only with the physical aspect of Nature a possible incompatibility between experimental science and theology, if any, would be in the physical aspect only. Herein lies the source of possible conflicts between experimental science and theology.  Metaphysics $m_{1}$ cannot contain metaphysics $m_{3}$ as a subset since the subject matter of metaphysics $m_{1}$ is the domain of the phenomenological context and thus regulative of only the purely physical. Therefore, a theistic worldview would be based on the metaphysics $m_{4}$, which is the union of $m_{1}$ and $m_{3}$, that is, $m_{4} = m_{1}\cup m_{3}$.  To insist of the exclusivity of metaphysics $m_{1}$ would correspond to a form of physicalism or materialism and thus the elimination of theology $t_{1}$. This is a form of reductionism, which violates the order of knowledge. The laws of experimental science are quite consistent with most theological propositions. It is in the study of unique historical events--say, in cosmological or biological evolution--where the conflict between science and religion may arise. Religion as a kind of knowledge is a synthetic or reducible kind of knowledge and is constituted by several of the autonomous or irreducible kinds of knowledge listed on Table 1.  For instance, the Christian faith is based essentially on the historicity of Jesus of Nazareth, his death, and his resurrection. Absent those historical events, there would be no Christian faith. Therefore, the Christian faith as a religion has an essential historical element in its constitution in addition to the theological propositions that underlie the supernatural aspect of the faith. Experimental science qua generalization of historical propositions has nothing whatsoever to say regarding a particular historical proposition that is not in the class of historical propositions that gave rise to the particular generalization. In other words, if $g_{1} \rightarrow (h_{1}, h_{2}\cdots, h_{n})$, where the parentheses denote a class of historical propositions, then one cannot conclude that the particular historical proposition $h_{n+1}$ is not possible owing to the generalization $g_{1}$, which is based on the class of historical propositions  $(h_{1}, h_{2}\cdots, h_{n})$. In particular, the results of experimental science cannot be used to disprove the possible existence of miracles as unique events. For instance, in 1931 Paul Dirac \cite{PAMD31} showed that the existence of magnetic monopoles leads to the quantization of electric charge, a very fundamental feature of Nature. Blas Cabrera \cite{BC82} presumably detected the hypothetical magnetic monopole. So far, however, this remains the only experimental detection and so it may be that magnetic monopoles do not exist at all or the event, whatever it was, is not reproducible. Therefore, in the order of being, the historicity of the magnetic monopole is on a par with that of a miracle owing to its historical uniqueness.

\begin{center}
\begin{tabular}{ |p{2cm}|p{2cm}|p{2cm}|  }
 \hline
 \multicolumn{3}{|c|}{Table 2: \textbf{Logical Implications} (T=True, F=False)  } \\
 \hline
 \textbf{p}  & \textbf{q} & \textbf{p} $\rightarrow$ \textbf{q}\\
 \hline
 \hline
 T &T&T\\
 \hline
 T&F&F   \\
 \hline
F&T&T\\
\hline
 F&F&T\\
 \hline
\end{tabular}
\end{center}

The regulative character of metaphysics for the experimental sciences and the other positive sciences, viz., formal logic, mathematics, and history, follows from the truth table of logical implications, Table 2.  Let the following implications be true for the experimental generalizations $(g_{1}, g_{2},\cdots, g_{20})$ and the metaphysical propositions ${m_{1}, m_{2}}$, viz.,  $(g_{1}, g_{2},\cdots,g_{10})  \rightarrow  m_{1}$ and $(g_{11}, g_{12}, \cdots, g_{20}) \rightarrow m_{2}$.  If $m_{1}$ and $m_{2}$ are incompatible, say $m_{1}$ is true and $m_{2}$ is false, then by Table 2 the set $(g_{11}, g_{12}, \cdots, g_{20})$ is false. Therefore, metaphysics is regulative of the positive sciences. Metaphysics is actually partially constitutive of theology, which is a stronger relation than metaphysics being merely regulative of theology. Note that the meaning of ``regulation" is such that an autonomous kind of knowledge A can be constitutive of another autonomous kind of knowledge B, if, in fact, A is not regulative of B. Thus in the implications for the theological propositions, M is constitutive of some propositions of T but is not regulative of T.

The term ``science" is equivocal since it can mean, (a) the method of arriving at generalization of historical propositions that constitute G, (b) the mass of information, accrued by such methods, and (c) the theories developed to summarize the data into laws of Nature. In addition, one needs to distinguish between the experimental sciences from the observational sciences, say, astronomy, paleontology, etc., and the historical sciences, say forensic science, cosmology, evolution of life on Earth, etc.  Here the term ``science" is defined by its subject matter, viz., the physical aspect of the whole of reality. Thus, science is the study of the ``real world around us" whereby the observer, us, is replaced from the outset from consideration \cite{ES67f}. This is accomplished here by considering the physical as the subject matter of science. The principle of objectivation, which together with the principle of understandability of Nature form the basis of the scientific method \cite{ES67e}, is accomplished here by considering the subject matter of science data that can be collected, in principle, by purely physical devices. Therefore, the laws of experimental science are generalizations of historical propositions, that is, experimental data, thus all physical laws are based on statistics. Note that consciousness and rationality are purely nonphysical, since purely physical devices cannot detect them. In addition, life cannot be reduced to the purely physical, so living beings are both physical and nonphysical. However, despite the difficulty of reducing life to the purely physical, Schr\"{o}dinger considers a genuinely physical, rather than nonphysical, not to say a supernatural, law in order to interpret life by the ordinary laws of physics \cite{ES67g}. Schr\"{o}dinger invokes the work of Max Planck dealing with the fundamental distinction between reversible and irreversible processes, viz., ``order-from-order" and ``order-from-disorder."  Accordingly, Schr\"{o}dinger's new principle for the understanding of life is nothing new to physics but the ``order-from-order," which is the same that governs large dynamical systems, say the motion of planets or clocks. Of course, it is not at all clear how the ``order-from-order" that already exists in living beings emerges from the dynamical ``order-from-order" governing purely physical systems.

\bibliography{}

\begin{thebibliography}{}
\bibitem{ET05} ``First: Thinking, alone, can never lead to any knowledge of external objects. Sense perception is the beginning of all research, and the truth of theoretical thought is given exclusively by its relation to the sum total of those experiences. Second: All elementary concepts are reducible to space-time concepts. Only such concepts occur in the `laws of nature.' In this sense, all scientific thought is `geometric.' A law of nature is expected to hold true without exceptions; it is given up as soon as one is convinced that one of its conclusions is incompatible with a single experimental fact. Third: The spatiotemporal laws are complete. This means, there is not a single law of nature that, in principle, could not be reduced to a law within the domain of space-time concepts. This principle implies, for instance, the conviction that psychic entities and relations can be reduced, in the last analysis, to processes of a physical and chemical nature within the nervous system. According to this principle, there are no nonphysical elements in the causal system of the processes of nature. In this sense, there is no room for `free will' within the framework of scientific thought, nor for an escape into `vitalism'." Albert Einstein, ``Physics, Philosophy, and Scientific Progress." Physics Today 58, no. 6 (2005), 46-48.
\bibitem{MA06} Moorad Alexanian, ``Set Theoretic Analysis of the Whole of Reality," Perspectives on Science and Christian Faith 58, no. 3 (2006), 254-55.
\bibitem{JM61} ``The crucial question for our age of culture is, thus, whether reality can be approached and known, not only `phenomenally' by science, but also `ontologically' by philosophy…...  In other words, being is the primary object of philosophy, as it is of human reason; and all notions worked out by philosophy are intelligible in terms of being, not of observation and measurement." Jacques Maritain, On the Use of Philosophy, (Princeton, New Jersey: Princeton University Press, 1961), 56-8.
\bibitem{CF04} Tim Crane and Katalin Farkas, eds.  Metaphysics: A Guide and Anthology, (New York: Oxford University Press, 2004), 610.
\bibitem{MA08} Moorad Alexanian, ``Can science make the `breath' of God part of its subject matter?"  Perspectives on Science and Christian Faith 60, no. 3 (2008), 207-8.
\bibitem{SN00a} Steven Nadler, ``Malebranche on Causation." In Malebranche, ed. Steven Nadler, (Cambridge: Cambridge University Press, 2000), 112-38.
\bibitem{SN97} Steven Nadler, ``Occasionalism and the Mind-Body Problem." In Studies in Seventeen-Century European Philosophy, ed. Michael A. Stewart, (Oxford: Clarendon Press, 1997), 75-95.
\bibitem{ES67} Erwin Schr\"{o}dinger, What is life? with Mind and Matter \& Autobiographical Sketches, (Cambridge, Great Britain: Cambridge University Press, 1967), 128.
\bibitem{JG92} Johann G\"{o}tschl, Erwin Schrödinger's World View: The Dynamics of Knowledge and Reality, (Dordrecht: Kluwer Academic Publishers, 1992).
\bibitem{PAMD29} P.A.M. Dirac, ``Quantum Mechanics of Many-Electron Systems.” Pro. Roy. Soc. London A123 (1929), 714-33.
\bibitem{ES83a} Erwin Schr\"{o}dinger, My View of the World, (Woodbridge, Connecticut: Ox Bow Press, 1983), 67.
\bibitem{ES83b} Ref. 11, 68.
\bibitem{ES83c} ``First, that the hypothesis of a material world as the cause of our wide area of common experience does nothing for our awareness of that shared character, that this awareness has to be achieved by thought just as much with this hypothesis as without it; secondly, I have stressed repeatedly, what neither can be nor needs to be proved, that this hypothetical causal connection  between the material world and our experience, in regard both to sense-perception and to volitions, differs toto genere from the casual relation which continues in practice, perfectly rightly, to play so important a part in science, even now that we have realized, with George Berkeley (b. 1685) and still more clearly with David Hume (b. 1711), that it is not really observable, not, that is, as a propter hoc but only as a post hoc.  The first of these considerations makes the hypothesis of the material world metaphysical, because there is nothing observable that corresponds to it; the second makes it mystical, because it requires the application of an empirically well-founded mutual relation between two objects (cause and effect) to pairs of objects of which only one (the sense-perception or volition) is ever really perceived or observed while the other (the material cause or material achievement) is merely an imaginative construct." Ref. 11, 93-4.
\bibitem{ES83d} Ref. 11, 94.
\bibitem{ES83e} Ref. 11, 37.
\bibitem{ES83f} Ref. 11, 101.
\bibitem{CSL78} C.S. Lewis, Mere Christianity, (New York: Macmillan Paperback edition, Twenty-eight Printing, 1978), 145.
\bibitem{CSL71a} C.S. Lewis, Miracles: A Preliminary Study,  (New York: The Macmillan Company, 1971), 25.
\bibitem{CSL71b} Ref. 18, 21.
\bibitem{ES56a} Erwin Schr\"{o}dinger, What is Life? \& Other Scientific Essays.  Garden City: Doubleday Anchor books, 1956), 182.
\bibitem{CSL71c} ``Human Reason and Morality have been mentioned not as instances of Miracle (at least, not the kind of Miracle you wanted to hear about) but as proofs of the Supernatural: not in order to show that Nature ever is invaded but that there is a possible invader." C. S. Lewis, Miracles: A Preliminary Study, (New York: The Macmillan Company, 1971), 44.
\bibitem{ES56b} Erwin Schr\"{o}dinger, What is Life? \& Other Scientific Essays, (Garden City: Doubleday Anchor books, 1956), 218.
\bibitem{CSL71d} ``When we are considering Man as evidence for the fact that this spatiotemporal Nature is not the only thing in existence, the important distinction is between that part of Man which belongs to this spatiotemporal Nature and that part which does not: or, if you prefer, between those phenomena of humanity which are rigidly interlocked with all other events in this space and time and those which have a certain independence. These two parts of man may rightly be called Natural and Supernatural." C.S. Lewis, Miracles: A Preliminary Study, (New York: The Macmillan Company, 1971), 175.
\bibitem {HB68a} H. Bondi, Cosmology.  London: Cambridge University Press, 1968), 65.
\bibitem{HB68b} Ref. 24, 100.
\bibitem{ES56c} ``We are thus facing the following strange situation. While all building stones for the world-picture are furnished by the senses qua organs of the mind, while the world-picture itself is and remains for everyone a construct of his mind and apart from it has no demonstrable existence, the mind itself remains a stranger in this picture, it has no place in it, it can nowhere be found in it. We are usually not aware of this.  We are so used, in our thoughts, to inserting the personality of a human being---that of an animal for that matter---into his body that we are amazed to learn and are doubtful and hesitant about believing that in reality it is not there." Erwin Schrödinger, What is Life? \& Other Scientific Essays, (Garden City- Doubleday Anchor books, 1956), 216.
\bibitem{SN00b} Steven Nadler, ``Malebranche on Causation." In Malebranche, ed. Steven Nadler, 112-38, (Cambridge: Cambridge University Press, 2000), 4.
\bibitem{ES83g} Erwin Schr\"{o}dinger, My View of the World, (Woodbridge, Connecticut: Ox Bow Press, 1983), 41.
\bibitem{ES83h} In inorganic beings, ``the essential and permanent element, the basis of identity and integrity, is the material, the matter, the inessential and mutable element being the form.  In the organic being the reverse is true; for its life, that is, its existence as an organic being, consists precisely in a constant change of matter while the form persists." Ref. 28, 42.
\bibitem{ES83j} Ref. 28, 101.
\bibitem{WL92} Werner Leinfellner, ``Schrödinger, the Self, and the Genes." In Erwin Schrödinger's World View: The Dynamics of Knowledge and Reality, ed. Johann Götschl, 87-98.  (Dordrecht: Kluwer Academic Publishers, 1992).
\bibitem{JM52} Jacques Maritain, The Range of Reason,  (New York: Charles Scribner’s Sons, 1952).
\bibitem{JM62} Jacques Maritain, An Introduction to Philosophy, (New York: Sheed and Ward, 1962).
\bibitem{WOM57} William Oliver Martin, The Order and Integration of Knowledge, (Ann Arbor: The University of Michigan Press, 1957).
\bibitem{MA07} Moorad Alexanian, ``Debate about science and religion continues," Physics Today (Letter) 60, no. 2 (2007), 10 \&12.
\bibitem{EG07} ``By extending the application of the term natural philosophy to the mathematical sciences, most notably mathematical physics, Newton may have begun a tradition that reached fruition in the nineteenth century when natural philosophy came to be called physics, and, often enough, science in general." Edward Grant, A History of Natural Philosophy, (New York: Cambridge University Press, 2007), 316.
\bibitem{ES67a} Erwin Schr\"{o}dinger, What is life? with Mind and Matter \& Autobiographical Sketches, (Cambridge, Great Britain: Cambridge University Press, 1967), 153.
\bibitem{ES67b} Ref. 37, 163.
\bibitem{JM52a} Jacques Maritain, The Range of Reason, (New York: Charles Scribner’s Sons, 1952), 4.
\bibitem{JM52b} Ref. 39, 6.
\bibitem{ES67c} Erwin Schr\"{o}dinger, What is life? with Mind and Matter \& Autobiographical Sketches, (Cambridge, Great Britain: Cambridge University Press, 1967), 93.
\bibitem{WJM89}  Walter John Moore, Schr\"{o}dinger: Life and Thought, (New York: Cambridge University Press, 1989), 225.
\bibitem{RL91} Rolf Landauer, ``Information is Physical,"  Physics Today 44, no. 5 (1991), 23.
\bibitem{ES67d} ``Let me briefly mention the notorious atheism of science which comes, of course, under the same heading. Science has to suffer this reproach again and again, but unjustly so. No personal god can form part of a world model that has only become accessible at the cost of removing everything personal from it. We know, when God is experienced, this is an event as real as an immediate sense perception or as one’s own personality. Like them he must be missing in the space-time picture. I do not find God anywhere in space and time—that is what the honest naturalist tells you. For this he incurs blame from him in whose catechism is written: God is spirit." Erwin Schr\"{o}dinger, What is life? with Mind and Matter \& Autobiographical Sketches, (Cambridge, Great Britain: Cambridge University Press, 1967), 138.
\bibitem{JM52c} ``But to posit such a `transubstantiation' between two entities which nevertheless retain their own being---for I remain what I am and the thing remains what it is while I know it---amounts  to saying that the process involves an immaterial becoming, an immaterial identification, and that knowledge is a dependent variable of immateriality. To know, therefore, consists of immaterially becoming another, insofar as it is another, aliud in quantum aliud." Jacques Maritain, The Range of Reason, (New York: Charles Scribner’s Sons, 1952), 12.
\bibitem{CSL71} C.S. Lewis, Miracles: A Preliminary Study, (New York: The Macmillan Company, 1971), 40.
\bibitem{MDW05} Margaret Dauler Wilson, Descartes, (New York: Taylor \& Francis, 2005), 159.
\bibitem{WOM57a} ``Now, whenever a science is considered from the standpoint of the mode studied, and not from the standpoint of the being that the mode is of, then that science is considered in its positive sense.  A positive science is one which is defined in abstraction from the metaphysical because it is concerned with a mode qua mode and not with a mode of being.  Experimental science is positive in this sense.  This does not mean that it is anti-metaphysical; it is simply non-metaphysical.  What is anti-metaphysical is positivism, which is the position that knowledge is limited to a `mode,' and that there can be no knowledge of a `mode of being' because there is no science of being qua being.  Experimental science is positive science, but it is not the same as positivism.  Positivism is a doctrine about it, and hence is not of it. A proposition about experimental science is not necessarily a proposition of it." William Oliver Martin, The Order and Integration of Knowledge, (Ann Arbor: The University of Michigan Press, 1957), 175-6.
\bibitem{TMS08} Tad M. Schmaltz, Descartes on Causation, (New York: Oxford University Press, 2008), 90.
\bibitem{TMS08a} Ref. 49, 4.
\bibitem{WOM57b} William Oliver Martin, The Order and Integration of Knowledge, (Ann Arbor: The University of Michigan Press, 1957), 319.
\bibitem{WOM57c} Ref. 51, 122.
\bibitem{ES67e} Erwin Schr\"{o}dinger, What is life? with Mind and Matter \& Autobiographical Sketches, (Cambridge, Great Britain: Cambridge University Press, 1967), 117.
\bibitem{PAMD31} P.A.M. Dirac, ``Quantised Singularities in the Electromagnetic Field," Proc. Roy. Soc.  London A 133 (1931), 60-72.
\bibitem{BC82} Blas Cabrera, ``First Results from a Superconductive Detector for Moving Magnetic Monopoles,"  Physical Review Letters 48 (1982), 1378–1381.
\bibitem{ES67f} Erwin Schr\"{o}dinger, What is life? with Mind and Matter \& Autobiographical Sketches, (Cambridge, Great Britain: Cambridge University Press, 1967), 161.
\bibitem{ES67g} Ref. 56, 80.
\end{thebibliography}

\end{document}